\documentclass[aps,pre,twocolumn,superscriptaddress,floatfix]{revtex4-1}
\usepackage{epsf,amsmath,amssymb,wasysym,color}
\newcommand{\be}{\begin{equation}}
\newcommand{\ee}{\end{equation}}

\begin{document}
\title{Spreading dynamics following bursty human activity patterns}
\author{Byungjoon Min} 
\author{K.-I.~Goh} 
\affiliation{Department of Physics, Korea University, Seoul 136-713, Korea}
\author{Alexei Vazquez} 
\affiliation{Department of Radiation Oncology, The Cancer Institute of New Jersey
and UMDNJ-Robert Wood Johnson Medical School, New Brunswick, NJ 08540, USA}
\date{\today}
\begin{abstract}
We study the susceptible-infected model with 
power-law waiting time distributions $P(\tau)\sim \tau^{-\alpha}$,
as a model of spreading dynamics under heterogeneous human activity patterns.
We found that the average number of new infections $n(t)$ 
at time $t$ decays as a power law in the long time limit, 
$n(t) \sim t^{-\beta}$, leading to extremely slow prevalence decay.
We also found that the exponent in the spreading dynamics, $\beta$,
is related to that in the waiting time distribution, $\alpha$, 
in a way depending on the interactions between agents but
is insensitive to the network topology.
These observations are well supported by both the theoretical
predictions and the long prevalence 
decay time in real social spreading phenomena.
Our results unify individual activity patterns with macroscopic
collective dynamics at the network level.
\end{abstract}
\maketitle
 
\section{Introduction}
Spreading dynamics represents many real social phenomena, such as
emerging epidemics and information flows, motivating its 
study for many years. 
Traditional epidemic modeling studies have assumed homogeneity 
in the timing of events and in the connectivity patterns of contacts
\cite{epidemic}. Recently modifications to classical models of spreading 
dynamics have been considered, to account for the complex connectivity
patterns of social networks \cite{sfnetworks} and the non-Poissonian 
nature of human activity patterns 
\cite{barabasi,surface,humandynamics,eckmann,caldarelli}. There have been 
intensive studies on the spreading dynamics in the presence of connectivity 
heterogeneity, in particular in networks with scale-free degree 
distribution $P(k)\sim k^{-\gamma}$ with $2<\gamma\le3$, that better 
represent the world wide web, co-authorship networks, and many other 
social networks \cite{vespignani,balcan,sanz}. On the other hand, 
the impact of time heterogeneity in human dynamics 
started to be recognized just recently. 
Various human activity patterns, ranging from tele-communications 
such as E-mail \cite{barabasi, eckmann}, surface mail \cite{surface},
web browsing \cite{humandynamics, web,radicchi}, instant messaging \cite{messaging},
and mobile phone calls \cite{candia}, to physical contacts probed by wireless devices \cite{wireless,cattuto}
or sexual contact survey \cite{sex}, have revealed
high heterogeneity regarding the timing of events, which is often 
described by a power-law waiting time distribution $P(\tau)\sim\tau^{-\alpha}$, 
where $\tau$ is the time interval between two consecutive
activities, a strong departure from traditional modeling assumption
\cite{humandynamics}. A few recent studies incorporating 
empirical heavy-tailed waiting time distribution in the spreading dynamics 
have demonstrated a significantly longer prevalence decay time 
than expected from the Poisson process \cite{alexei,iribarren}.

On the individual level, priority-queue models mimicking human decision 
process were introduced to reproduce power-law waiting time distributions 
\cite{barabasi,humandynamics,caldarelli}. In these models, prioritization 
of tasks is considered a major cause of heavy-tailed nature of human 
dynamics. These models were extended to include interactions between two 
or more agents inevitable in social system \cite{and,pqn,cho}.
While these models are successful in reproducing 
power-law waiting time distributions, it is still veiled how the activity 
modeling at the individual level translates into the collective 
phenomena at the population or network level. So, here we study spreading 
dynamics as a representative example of macroscopic phenomena in social 
system using the susceptible-infected (SI) model under heterogeneous 
activity patterns.

\begin{figure*}
\centerline{\epsfxsize=0.66666\linewidth \epsfbox{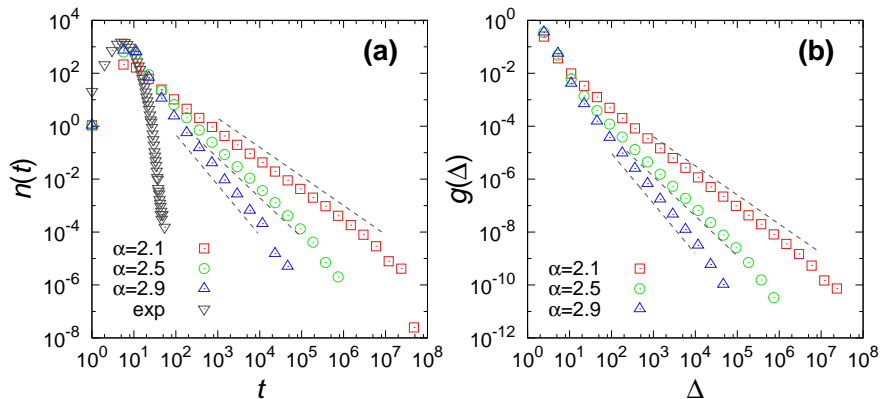}}
\caption{
(Color online) 
(a) The average number of new infections $n(t)$ of the SI model
with uncorrelated identical power-law $P(\tau)$
with exponent $\alpha=2.1$ ($\Box$), $2.5$ ($\circ$), $2.9$ ($\triangle$), 
and with an exponential $P(\tau)$ ($\triangledown$), respectively.
All $P(\tau)$ have the same mean waiting time $\langle\tau\rangle=2$.
It decays with the exponent $\approx \alpha-1$ for power-law $P(\tau)$,
whereas it decays exponentially for exponential $P(\tau)$.
(b) The generation time distribution $g(\Delta)$ for power-law $P(\tau)$.
It decays as a power law with the same exponent as $n(t)$.
Simulations were performed on a scale-free network with $\gamma=3$
and size $N=10^{4}$, and averaged over $10^{8}$ independent runs.
Initial infected node is selected to be the hub.
In both panels, dotted lines have slopes $-1.1$, $-1.5$, and $-1.9$,
drawn as a guide. 
Data are binned logarithmically.
}
\end{figure*}

\section{SI model with power-law waiting time distributions}
\subsection{General theory}
To start understanding 
the impact of power-law waiting time distribution,
we first turn to a general theory of irreversible spreading processes 
in a social network. 
We assume the network has a tree-like structure and characterize the
timing of infection transmissions by
the generation time distribution $g(\Delta)$, where the 
generation time $\Delta$ is defined as the time interval between the 
infection of an agent in the social network (primary case) and the 
transmission of the infection to a neighbor agent (secondary case). In 
this case, an outbreak starting from a single infected individual at 
$t=0$ results in the average number of new infections
$n(t)=\sum_{d=1}^{D} z_d g^{\star d}(t)$ at time $t$ \cite{alexei2},
where $z_d$ is the average number of individuals $d$ contacts away from 
the first infected node, $D$ is the maximum of $d$, and $g^{\star d}(t)$ 
is the $d$-th order convolution of $g(\Delta)$ [$ g^{\star 1}(t)=g(t)$ and 
$g^{\star d}(t)=\int_0^{t} dt' g(t') g^{\star d-1} (t-t')$], 
representing the probability density function of the sum of $d$ generation 
times. For the cases where $g(\Delta) \sim \Delta ^{-\beta}$ with  $1<\beta<2$, 
in the limit $d \gg 1$ one obtains
$g^{\star d}(t)\sim \mathcal{L}_{\beta-1}(t/t_d)/t_d$, 
where $t_d=\Delta_0 d^{\frac{1}{\beta-1}}$, $\Delta_0$ is some characteristic time scale and 
$\mathcal{L}_\mu(x)$ represents the stable (L{\'e}vy) distribution with 
exponent $\mu$ concentrated in [0,$\infty$), characterized by
the asymptotic behavior $\mathcal{L}_\mu(x) 
\sim x^{-(1+\mu)}$ for $x \gg 1$ \cite{feller}. From this it follows that, 
independently of the network structure,
$g^{\star d}(t) \sim t^{-\beta}$ when $t\to\infty$ and
\be\label{ntbeta}
n(t)\sim t^{-\beta}.
\ee
Therefore, if the generation time distribution decays as a power law with exponent $\beta$, then the long
time dynamics of the spreading process will be characterized by the same power-law decay.
In general, when the generation time distribution is heavy-tailed but not power-law, we expect that
the long-time spreading dynamics decay with the same tail as the generation time distribution by the above theory.

\subsection{SI model with uncorrelated power-law activity patterns}
The generation time distribution, in turn, can be determined from
the waiting time distribution and the type of interactions 
between the social agents.
As the simplest scenario of spreading process, let us first consider 
the case when the acitivities of individuals in the network are
completely uncorrelated and so are the timings of consecutive interactions.
In such a case, the time series of interactions can be modeled by 
a simple renewal process \cite{feller} with waiting time distribution $P(\tau)$. 
Specifically, let us consider the scenario where the agent A is infected 
at time $t_{A}$, B, who is connected to A, still susceptible at $t_{A}$, 
and B is infected at time $t_{B}$ after interaction with agent A.
From the perspective of B, the time A was infected is random and therefore,
the generation time $t_{B}-t_{A}$ is the residual waiting time, 
the time interval between a randomly selected time and the time A and B 
will perform their next interaction. 

For the waiting time distribution with finite mean $\langle\tau\rangle$, 
the residual waiting time probability density function is related to the 
waiting time probability density function, leading to 
$g_{uncorr}(\Delta)=\frac{1}{\langle\tau\rangle} 
\int_{\Delta}^{\infty}P(\tau)d\tau$ \cite{feller}. Therefore, for the 
uncorrelated activity patterns with $P(\tau) \sim \tau^{-\alpha}$, 
$2<\alpha<3$, we have 
$g_{uncorr}(\Delta)\sim\Delta^{-(\alpha-1)}$ and $n_{uncorr}(t)\sim 
t^{-(\alpha-1)}$. On the other hand, when $1<\alpha<2$, $P(\tau)$ does not 
have a finite mean so the generation time distribution cannot be obtained 
in the same way. In this case the residual time distribution is non 
stationary and, in the long time limit, it approaches the limit 
distribution $g_{uncorr}(\Delta,t) \rightarrow (1/t) p_\alpha(\Delta/t)$ 
\cite{feller}, where $0\leq\Delta\leq t$ and $p_\alpha(x) = 
\sin\pi\alpha/[\pi x^\alpha(1+x)]$. Furthermore, in this case the 
convolutions of $g_{uncorr}(\Delta,t)$ are of the form $g^{\star 
d}_{uncorr}(\Delta,t) \rightarrow (1/t)p_\alpha^{\star d}(\Delta/t)$. Substituting the 
latter result into $n(t)=\sum_{d=1}^{D} z_d g^{\star d}(t)$ we obtain, in 
the long time limit, $n(t)\approx t^{-1} \sum_{d=1}^D z_d p_\alpha^{\star 
d}(1)$. Putting all together we thus obtain $g_{uncorr}(\Delta)\sim 
\Delta^{-\beta_{uncorr}}$ for $2<\alpha<3$ and $n_{uncorr}(t)\sim 
t^{-\beta_{uncorr}}$ for $1<\alpha<3$ with

\be\label{betaunc}
\beta_{uncorr} = 
\left\{
\begin{array}{ll}
1, & 1<\alpha\leq2\\
\alpha-1, & 2<\alpha<3
\end{array}
\right.
\ee

To test the predicted power-law decay of $n(t)$, we perform
the following numerical simulations.
Initially all agents are susceptible except for a single infected node.
Then infected agents infect connected susceptible agents  
following a power-law waiting time distribution
with exponent $\alpha$ until all agents are infected.
For $2<\alpha<3$, $n(t)$ is found to decay as a power law with the exponent 
$\alpha-1$ for large $t$ as predicted by the theory (Fig.~1).
For comparison, we also performed the same SI dynamics with 
exponential (Poisson-type) waiting time distribution with the same
mean waiting time $\langle\tau\rangle=2$ [Fig.~1(a), $\triangledown$].
In this case, $n(t)$ decays exponentially fast, in stark contrast to
the power-law cases. Specifically, the effective duration $T$ of the epidemic
process, given by the expected infection time of an individual after 
the outbreak, $T=\sum_{t=0}^{\infty}tn(t)/N$, is measured to be $T_{exp}\approx5$ 
for the exponential $P(\tau)$, whereas it can become orders of magnitude
longer as $T\approx4\times10^5$ for the power-law $P(\tau)$ with $\alpha=2.1$.
Therefore, the power-law waiting time distribution indeed impacts
the long-time dynamics of the spreading process significantly.

\begin{figure}[t]
\centerline{\epsfxsize=.9\linewidth \epsfbox{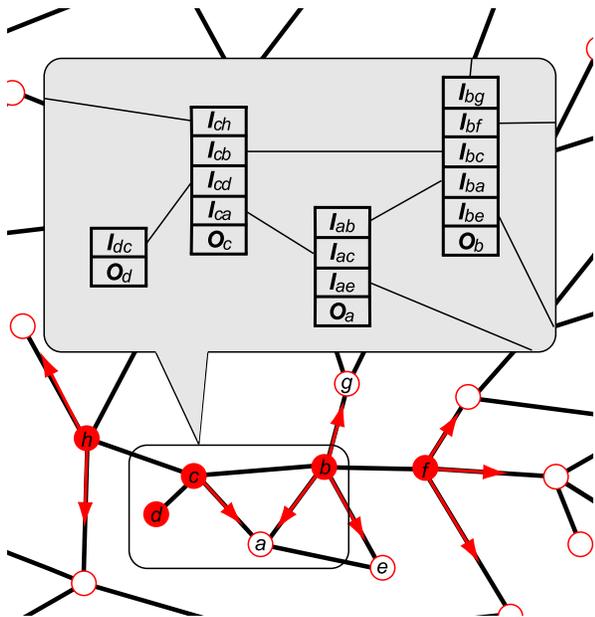}}
\caption{
(Color online) Schematic illustration of the SI model on the PQN.
Filled (open) nodes denote infected (susceptible).
The arrows from an infected to a susceptible node indicate
potential subsequent infection channels.
}
\end{figure}
\begin{figure*}
\centerline{\epsfxsize=\linewidth \epsfbox{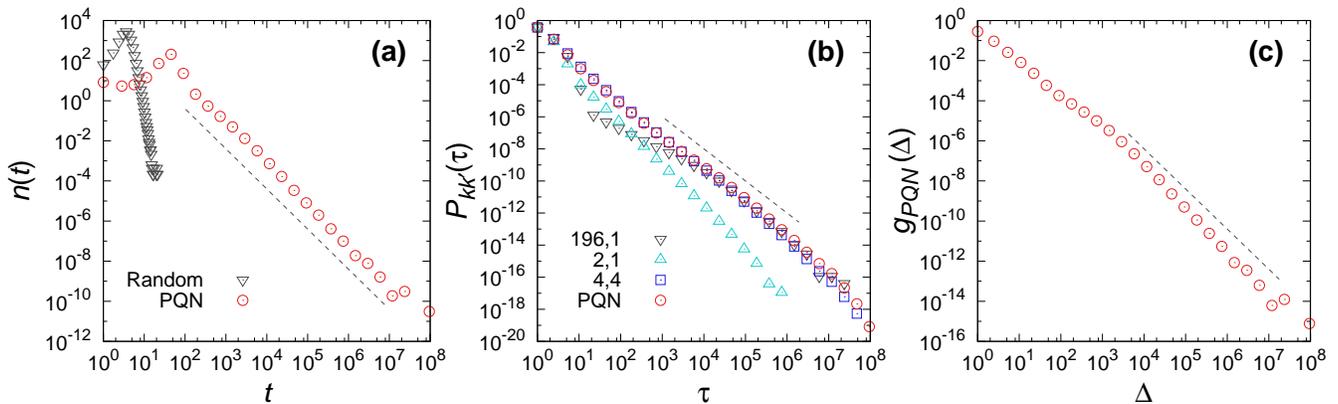}}
\caption{
(Color online) (a) The average number of new infection $n(t)$ 
of the SI model on the PQN ($\circ$) and on the random execution
queue network ($\triangledown$).
After the initial increase, $n(t)$ after the peak decays with a power-law tail 
with exponent $\beta\approx2$ for the PQN ($\circ$) and exponentially for
the random execution case ($\triangledown$).
(b) The waiting time distribution of the whole network, $P_{PQN}(\tau)$, and the 
waiting time distributions of links connecting nodes with degree $k$ and $k'$,
$P_{kk'}(\tau)$, of the same process for the PQN.
Both $P_{PQN}(\tau)$ and $P_{kk'}(\tau)$ decay as a power law,
yet with different exponents dependent on the local 
topological position, $\alpha_{PQN}\approx 2$, $\alpha_{196,1}\approx 1$, 
$\alpha_{2,1} \approx 3$, and $\alpha_{4,4}\approx 2$.
(c) The generation time distribution $g_{PQN}(\Delta)$ of the same process 
for the PQN. It decays as a power law with the same exponent as $n_{PQN}(t)$.
In all panels, dotted lines have slope $-2$, drawn as a guide.
Simulations were performed on a scale-free network with $\gamma=3$ and
size $N=10^{4}$, and averaged over $10^{3}$ different initial conditions.
Data are binned logarithmically.
}
\end{figure*}

\section{SI model on the PQN}
In a more realistic scenario,
the multi-agent dynamics in a social network
based on human decision process can be modeled  
by the priority-queue network (PQN) with interactions (Fig.~2).
We consider the SI model on PQN with {\sc or}-type interactions, which
can model the spreading of infectious entity through mutual communications
or contacts between individuals that can be initiated primarily 
by any one of the individuals, such as talking over the phone
or face-to-face encounter.
Specific procedures of numerical simulations are as follows.
Each node in a network is a queue with two kinds of tasks, 
interacting ($I$) and 
non-interacting ($O$) task \cite{and,pqn}.
A node $i$ with degree $k_i$ is a queue with fixed length $k_i+1$,
having $k_i$ $I$ tasks and one $O$ task.
The $I$ task of the node $i$ paired with the node $j$ 
through the link between them is denoted by $I_{ij}$ 
and the $O$ task of the node $i$ is denoted by $O_{i}$ (Fig.~2).
We start with all nodes being susceptible, except for one
infected node selected to be the hub.
Initially a random priority value from uniform distribution 
is assigned to each task.
Each step, we choose a node $i$ randomly 
and the highest priority task in the queue is executed.
At this step, if the highest priority task in the queue
is $O_i$ task, then it is executed alone.
However if it is $I_{ij}$ task, then not only $I_{ij}$ task
but also its conjugate task $I_{ji}$ are simultaneously executed.
If an infected and a susceptible node perform the $I$ task between them, 
the ``disease'' is transmitted and the susceptible node becomes infected.
Then all executed tasks are replaced by new tasks each with a random priority.
This infection process proceeds until all nodes are infected.
We have checked that different choices of the initial infected node other
than the hub do not affect the long-time dynamics of spreading processes.

We first consider the PQN with the scale-free degree distribution 
with $\gamma=3$.
In the long time limit, $n(t)$ is found to decay as a power law 
with exponent approximately $2$ [Fig.~3(a),~$\circ$].
The power-law decay of $n(t)$ is much slower than
the exponential decay in the SI model on the random execution queue network
[Fig.~3(a),~$\triangledown$].
Indeed such a longer prevalence time than predicted by the Poisson process 
was observed in real spreading dynamics controlled by human activity patterns, 
such as an E-mail virus outbreak \cite{vespignani,alexei}.

To understand the $n(t)\sim t^{-2}$ behavior observed for the PQN, 
we turn to the waiting time distribution of PQN, through which the generation
time distribution can be obtained.
It has been shown that there is a local variation of the waiting time
distribution exponent $\alpha$ in the PQN \cite{pqn}, which renders
the identification of network-level exponent $\alpha_{PQN}$ nontrivial.
Moreover, the highly coupled and non-Markovian nature of
PQN dynamics makes its analytic understanding currently a formidable challenge,
even for the simplest case with $N=2$ \cite{and,pqn}.
Lacking exact solutions, however, we could infer its key long-time dynamic
characteristics from the following statistical reasoning:
We first observe that different $I$ tasks (links) 
manifest different exponents $P_{kk^\prime}(\tau)\sim \tau^{-\alpha_{kk^\prime}}$, 
parametrized by the degrees $k$ and $k^\prime$ of the associated nodes.
For example, for the $I$ tasks between the hub and a node 
with $k=1$, $\alpha_{k_{hub},1}=1$, whereas $\alpha_{2,1} \approx 3$ [Fig.~3(b)].
Putting the contribution of all tasks together we obtain
$P_{PQN}(\tau)=\sum_{kk^\prime} e_{kk\prime} P_{kk^\prime}(\tau)$, where $e_{kk\prime}$ is the fraction of
links connecting nodes with degree $k$ and $k^\prime$. The smallest the exponent $\alpha_{kk^\prime}$ the biggest is
its contribution to the tail of $P_{PQN}(\tau)$, except for the fact that links with $\alpha_{kk^\prime}<2$ cannot contribute statistically
because of its strong peak of $P_{kk^\prime}(\tau)$ at $\tau \approx 1$. The net result is that links with $\alpha_{kk^\prime}\approx 2$
contribute the most, resulting in [Fig.~3(b), $\circ$],
\begin{equation} P_{PQN}(\tau)\sim \tau^{-\alpha_{PQN}}, \quad
\alpha_{PQN}= 2.
\end{equation}

\begin{figure}
\centerline{\epsfxsize=.9\linewidth \epsfbox{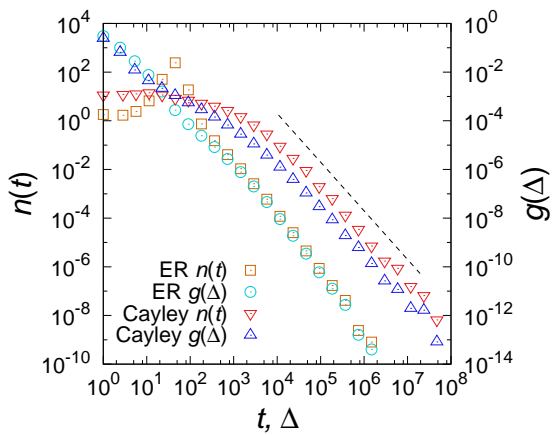}}
\caption{
(Color online) 
$n(t)$ (left scale) and $g(\Delta)$ (right scale) of the SI model on the Erd{\H{o}}s-R{\'e}nyi (ER) network 
with mean degree $\langle k \rangle = 4$ and size $N=10^{4}$ and
Cayley tree with branching number $k=4$ with $6$ generations ($N=1,456$).
In each case, $n(t)$ and $g(\Delta)$ show power-law tails with the same exponent.
Dotted line has slope $-2$, drawn as a guide.
Data are binned logarithmically.
}
\end{figure}

To proceed with relating the generation time distribution
with the waiting time distribution, we note that
the interaction timings between connected pairs are not completely 
independent in the PQN.
To see this explicitly, let us consider three linearly connected nodes 
($\mathrm B - \mathrm A - \mathrm C$) in the PQN. 
While node $\mathrm A$ interacts with node $\mathrm B$ ($I_{AB}$ task is being executed),
node $\mathrm A$ cannot interact with node $\mathrm C$ ($I_{AC}$ task has to wait).
So the waiting time of $I_{AC}$ begins at the execution of $I_{AB}$ task.
If the node $\mathrm B$ has been infected, the generation time 
of infection of $\mathrm C$ from $\mathrm A$ is the same as the waiting time.
Therefore, in the PQN the generation time distribution 
$g_{PQN}(\Delta)$ has the same power-law exponent as the waiting time 
distribution $P_{PQN}(\tau)$ [Fig.~3(c)], which in turn, 
according to our general theory, leads to $n_{PQN}(t)$ 
decaying with the same power law as $P_{PQN}(\tau)$, that is
\begin{eqnarray}\label{gPQN}
\beta_{PQN}&=&\alpha_{PQN}=2,
\end{eqnarray}
consistent with our numerical findings.

Finally, we examine the influence of the network structure 
on the spreading dynamics by the SI model over the PQN 
on Erd{\H{o}}s-R{\'e}nyi network and Cayley tree.
On both network structures, $n(t)$ is found to decay as a power law 
also with $\beta = \alpha$, $n(t)\sim t^{-\alpha}$ (Fig.~4).
Since every node except the root in Cayley tree is topologically 
indistinguishable, all $I$ tasks show the same $P(\tau)$,
which decays as a power law with $\alpha\approx2$ for degree $k\ge3$.
These results suggest that the network structure  
exerts minor influence on the spreading dynamics in the long time limit
and the heterogeneous activity pattern is the
major determinant of the long time dynamics.

\section{Summary and Discussion}
To conclude, we studied the SI model following heterogeneous activity 
patterns in time. 
The analytic prediction indicates that the
power-law waiting time distribution $P(\tau)$ leads to
the power-law decaying new infection number $n(t)$ in the long time limit, 
with the exponent determined through the generation time distribution $g(\Delta)$.
These theoretical predictions are in good agreement with 
numerical simulation results for
two different scenarios of spreading process by the SI model on 
the priority-queue networks and that under uncorrelated 
power-law waiting time distributions,
respectively.
In this study, we have demonstrated how the macroscopic phenomenon 
of extremely long prevalence 
time in spreading dynamics emerges from individual-level 
activity model with interactions.
Our results indicate that the heterogeneous individual activity patterns 
significantly impact social dynamics and 
therefore it is an essential
factor in modeling both the individual and the network level dynamics,
such as the spreading of human infectious diseases \cite{epidemic}
and that of mobile phone viruses \cite{wang},
which is expected to be a major problem in the near future.

\begin{acknowledgments}
This work was supported by the Korea Research Foundation Grant
funded by the Korean Government (MOEHRD, Basic Research Promotion
Fund) (KRF-2008-314-C00377).
B.M.\ acknowledges the support from the Seoul Scholarship Foundation.
\end{acknowledgments}

\end{document}